\documentclass{article}
\usepackage{hip-artc}
\volnumber{} \issuenumber{} \edyear{2003}                                
\frompage{000} \topage{000}                                              
\recrevdate{15 May 2003}                                                 
\def\rightmark{Kaons in nuclear medium}
\def\leftmark{C.L.\ Korpa and M.F.M.\ Lutz}

\title{Kaon and antikaon properties
in cold nuclear medium}
\authors{
{C.L.\ Korpa$^1$ and M.F.M.\ Lutz$^{2}$%
\index{One, A.}
\index{Two, A.}
}\\[2.812mm]
{\normalsize
\hspace*{-8pt}$^1$ Dept.\ of Theoretical Physics, University of P\'ecs\\
7624 P\'ecs, Hungary\\[0.2ex]
\hspace*{-8pt}$^2$ Gesellschaft f\"ur Schwerionenforschung (GSI)\\
64291 Darmstadt, Germany
}}

\abstract{We present results of a self-consistent calculation for
the kaon and antikaon spectral functions in cold nuclear matter,
using as input the kaon-nucleon and antikaon-nucleon scattering
amplitudes of the vacuum. We investigate the effect of in-medium
pion dressing on the antikaon-nucleon scattering amplitudes and
antikaon spectral function. We find the influence of pion dressing
to be minor on the antikaon spectral function and limited on the
hyperon resonances causing only a small additional broadening. An
exception is the $\Sigma(1690)$.  At nuclear saturation density an
attractive mass shift of about 20 MeV and width of about 130 MeV
is obtained. The kaon shows a repulsive mass increase of $36\,$MeV
and a small width of the quasiparticle peak at saturation density.
}


\makeindex
\begin{document}

\maketitle
\setcounter{page}{1}

\section{Introduction}\label{intro}
Interest in kaon properties in nuclear medium increased greatly after
it was suggested \cite{kaplan86} that antikaons may condense in the
interior of neutron stars. Describing kaonic atoms \cite{friedmann94}
and strangeness production in heavy-ion reactions \cite{laue99}
also requires knowing the in-medium spectral function of kaons and
antikaons.

The theoretical work on kaon properties in the nuclear medium in the
last decade was extensive
\cite{njl-lutz,brown94,brown96,koch94,waas96,waas97,ohnishi97,lutz98,ramos00,%
tolos01,tolos02,lutz02,korpa03}. For the antikaons
(K$^-$ and $\bar{\rm{K}}^0$) the results
confirmed a considerable softening expected on the basis of K-matrix
analysis \cite{martin81}, although quantitative differences remain
in the shape of the spectral function and its average shift towards
smaller energy. The kaons (K$^+$ and ${\rm{K}}^0$), on the other hand,
show a positive mass shift with little broadening of the spectral
function.

\section{Formalism}\label{formalism}
The computational scheme consists of a generalization of the
self-consistent approach presented in ref.~\cite{lutz02}. It uses
the in-medium kaon-nucleon and antikaon-nucleon scattering
amplitude, which take into account the medium's influence on the
particles' propagation. This is achieved by solving the in-medium
Bethe-Salpeter (BS) equation which contains the in-medium
propagators, but is based on the vacuum BS kernel. Inclusion of
relevant meson-hyperon channels leads to the matrix BS equation in
vacuum, written in compact way as:
\begin{equation}
T=K+K\cdot G\cdot T,
\end{equation}
where $T$ is the scattering amplitude, $K$ the kernel and $G$ denotes
the two-particle propagator. In the medium the same equation takes the
form:
\begin{equation}
{\mathcal T}={\mathcal K}+{\mathcal K}\cdot{\mathcal G}\cdot{\mathcal T}.
\label{hatt}
\end{equation}
Under the assumption that the kernels are the same, ${\mathcal K}=K$,
eq.~(\ref{hatt}) can be rewritten:
\begin{equation}
{\mathcal T}=T+T\cdot \Delta G\cdot{\mathcal T},\quad \Delta G\equiv
{\mathcal G}-G.
\label{hattnew}
\end{equation}

More explicitly, the (pseudo)scalar-meson--spin-1/2 baryon on-shell
scattering amplitude is written as
\begin{eqnarray}
&&\langle M(\bar q)\,B(\bar p)|\,T\,| M (q)\,B(p) \rangle
=(2\pi)^4\,\delta^4(q+p-\bar q-\bar p)\,
\nonumber\\
&& \qquad \qquad   \qquad \qquad  \qquad \qquad \times \,\bar u_B(\bar p)\,
T_{MB \rightarrow MB}(\bar q,\bar p ; q,p)\,u_B(p) \,,
\label{on-shell-scattering}
\end{eqnarray}
and satisfies the BS equation:
\begin{eqnarray}
T(\bar k ,k ;w ) &=& K(\bar k ,k ;w )
+\int \frac{d^4l}{(2\pi)^4}\,K(\bar k , l;w )\, G(l;w)\,T(l,k;w )\;,
\nonumber\\
G(l;w)&=&-i\,S_N({\textstyle
{1\over 2}}\,w+l)\,D_{\bar{K} }({\textstyle {1\over 2}}\,w-l) \,,
\nonumber\\
w &=& p+q = \bar p+\bar q\,,
\quad k= \frac{1}{2}\,(p-q)\,,\quad
\bar k =\frac{1}{2}\,(\bar p-\bar q)\,,
\label{BS-eq}
\end{eqnarray}
containing the free space baryon propagator
$S_B(p)=1/(p\!\!\!/-m_B+i\,\epsilon)$ and the meson propagator
$D_{M}(q)=1/(q^2-m_M^2+i\,\epsilon)$. The in-medium
eq.~(\ref{hatt}) has the same form, apart from the fact that the
4-point Green function ${\mathcal T(\bar k,k;w,u)}$ and the
2-particle propagator depend also on the 4-velocity $u_\mu$
characterizing the nuclear matter frame. For nuclear matter moving
with a velocity $\vec v$:
\begin{equation}
u_\mu =\left(\frac{1}{\sqrt{1-\vec v\,^2/c^2}},\frac{\vec v/c}
{\sqrt{1-\vec v\,^2/c^2}}\right)
\;, \quad u^2 =1\,.
\end{equation}

For the calculation of kaon and antikaon self energies in the
nuclear medium we need the kaon-nucleon and antikaon-nucleon
in-medium scattering amplitudes (more precisely, the 4-point Green
functions). In the case of antikaons a successful description of
free-space scattering data requires the inclusion of various
inelastic meson-baryon channels $X={\pi \Lambda\,, \pi \Sigma\,,
\eta \Lambda\,, \eta \,\Sigma\,, K\,\Xi }$ with strangeness $-1$
(see e.g. ref.~\cite{lutz02a}). As one sees from
eq.~(\ref{hattnew}) the need for the explicit inclusion of such
channels for the in-medium computation depends on size of the
modification of the meson-baryon propagator but also on the size
of the relevant transition amplitude $T_{\bar K N \to X}$. Note
that the contribution of such channels involving the free
propagators is already taken care of in $T_{\bar K N \to \bar K
N}$. Since we do not consider nuclear densities high enough for
hyperon condensation, only meson propagator modification should be
considered. It has been claimed in ref.~\cite{tolos02} that the
medium-modified pion influences significantly the
antikaon properties. It is known that the pion spectral function
changes considerably in the medium, however its influence on the
antikaon-nucleon channel depends on the $T_{\pi Y\rightarrow \bar
K N}$ transition amplitudes. The higher mass of the $\eta$-meson
leads to a smaller expected effect since already the influence of
the $\eta Y$ channels are of minor for the low-energy
antikaons-nucleon scattering. Motivated by these considerations we
do not include $\eta Y$ channels in our calculation, dropping as
well the kaon-cascade channels.

We work with definite isospin amplitudes and thus consider separately
the $I=0$ and $I=1$ cases. We have three coupled channels for $I=1$:
$\bar K N, \pi \Lambda(1116), \pi \Sigma(1195)$, and only two
for $I=0$: $\bar K N$ and $\pi \Sigma(1195)$. The meson-baryon
propagator matrix is diagonal and for $I=1$ has the form:
$\Delta G^{(0)}=\textrm{diag}(\Delta G_{\bar{K}N},
\Delta G_{\pi\Sigma},\Delta G_{\pi\Lambda})$, while for $I=0$
only the first two entries are present. Since we consider only
isospin-symmetric medium, the propagators in the two channels are
identical. When solving the matrix BS equation we use the fact that
the pion dressing is performed based on the pion-nucleon scattering
amplitude, which means that it is independent of the antikaon dressing.

We illustrate the method on the case of the $I=0$ channel. For simplicity
we denote the channel indes $\bar K N$ by ``1'' and $\pi\Sigma$ by ``2''.
Then writing out the [11] component of the matrix BS equation gives:
\begin{equation}
{\mathcal T}_{11}=T_{11}+T_{11}\,\Delta G_1\,{\mathcal T}_{11}
+T_{12}\,\Delta G_2\,{\mathcal T}_{21},
\label{comp11}
\end{equation}
where $\Delta G_1$ denotes $\Delta G_{\bar K N}$ and
$\Delta G_2$ means $\Delta G_{\pi\Sigma}$. Similarly, the [21]
component allows us to solve for the ${\mathcal T}_{21}$ in
terms of ${\mathcal T}_{11}$:
\begin{equation}
{\mathcal T}_{21}=(1-T_{22}\,\Delta G_2)^{-1}\left[
T_{21}+T_{21}\,\Delta G_1\,{\mathcal T}_{11}\right].
\label{comp21}
\end{equation}
Substituting eq.~(\ref{comp21}) into (\ref{comp11}) we get:
\begin{equation}
{\mathcal T}_{11}=\hat T_{11}+\hat T_{11}\,\Delta G_1\,{\mathcal T}_{11},
\label{comp11n}
\end{equation}
where we introduced
\begin{equation}
\hat T_{11}\equiv T_{11}+T_{12}\,\Delta G_2\,(1-T_{22}\Delta G_2)^{-1}\,T_{21}.
\label{that11}
\end{equation}
The form of eq.~(\ref{comp11n}) is the same as for the case with no pion
dressing, only the vacuum scattering amplitude $T_{11}$ has been replaced
by the right-hand side of (\ref{that11}). An analogous, though somewhat more
involved scheme can be used for the $I=1$ channel. This means we have to calculate the
effect of pion dressing only once and then perform the iterative, self-consistent
calculation with antikaon dressing using the obtained input which, however, has
the full in-medium structure (containing mixing of partial waves) and depends
on the energy and magnitude of the 3-momentum.

\section{Results}\label{results}
\begin{figure}[htb]
\vspace*{-0.1cm}
                 \insertplot{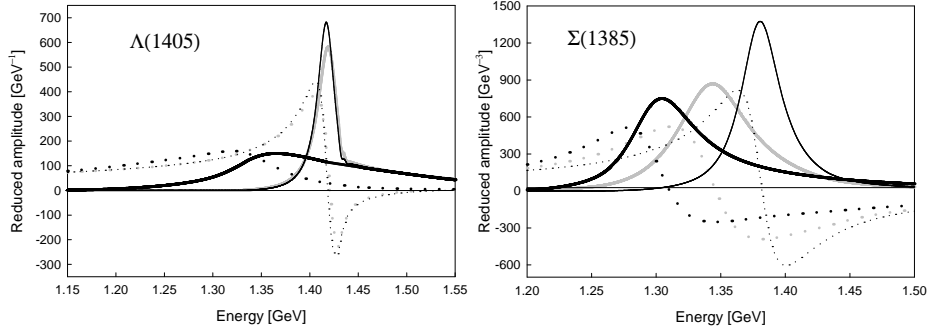}

\vspace*{-3.4cm}
\caption[]{The reduced scattering amplitudes corresponding to the
$\Lambda(1405)$ and $\Sigma(1385)$ at zero 3-momentum in nuclear
medium of density $\rho=0.17\,$fm$^{-3}$. Full lines show the imaginary
part, dotted lines the real part. The black lines show the effect of both
dressed pion and antikaon, the gray lines of dressed pion only. The thin black
lines show the vacuum amplitudes for comparison.
}
\label{fig1}
\end{figure}

For the dressed pion propagator we use the results of a
self-consistent calculation \cite{korpa03a} using as an input the
vacuum pion-nucleon scattering amplitudes, as well as the
relativistic delta-hole model of ref.~\cite{lutz03}. The main
difference between the two results is that the second one provides
somewhat more attraction, leading to a more pronounced softening
of the pion spectrum in the medium. As explained in the previous
section we first perform a calculation taking into account the
pion dressing and then use that as input for the self-consistent
computation of the antikaon spectral function. A hint for the
importance of pion dressing is given already by these input
amplitudes, i.e.\ their departure from the vacuum ones. In
fig.~\ref{fig1} we show the amplitudes corresponding to the
$\Lambda(1405)$ and the $\Sigma(1385)$ as obtained after including
the pion dressing (gray lines) and both pion and antikaon dressing
(black lines). The thin black lines show the vacuum amplitudes.
For fig.~\ref{fig1} we used the pion spectral function as obtained
in a relativistic delta-hole model \cite{lutz03}, which produces
slightly more pronounced effect than the pion spectral function
from a self-consistent calculation of ref.~\cite{korpa03a}. As
compared to our previous results \cite{lutz02} the $\Lambda(1405)$
resonance is basically unchanged only for the $\Sigma(1385)$ a
somewhat larger attractive mass shift is obtained. Further results
not shown in the figures concern the $\Sigma(1690)$ resonance for
which no significant effect was found previously
ref.~\cite{lutz02}. At saturation density pion dressing shifts the
resonance mass down by $20$ MeV and leads to a sizeable increase
of its decay width ($\simeq 130$ MeV).

\begin{figure}[htb]
\vspace*{-0.2cm}
                 \insertplot{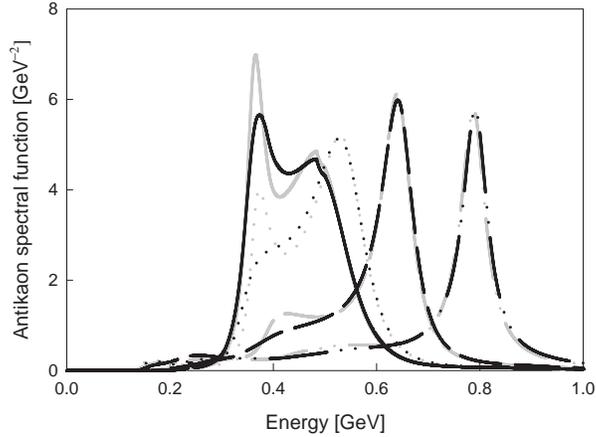}

\vspace*{-1.4cm}
\caption[]{The antikaon spectral function for different momenta:
0 (solid line), $200\,$MeV (dotted line), $400\,$MeV (dashed line),
and $600\,$MeV (dash-dot-dot line). Black lines show the result
of both dressed pion and antikaon, gray lines previous results
\cite{lutz02} without pion dressing, in the nuclear matter at
saturation density.
}
\label{fig2}
\end{figure}

Based on the observation that the antikaon properties in the
nuclear medium are determined mostly by the properties of the
s-wave $\Lambda(1405)$ resonance and to a smaller effect by the p-
and d-wave resonances, we expect the inclusion of pion dressing
not to produce dramatic changes in the antikaon spectral function.
These expectations are confirmed by the full self-consistent
calculation, whose results for the antikaon spectral function are
shown in fig.~\ref{fig2}. We observe that the pion dressing has a
small effect on the antikaon spectral function, limited to momenta
below $400\,$MeV. We do not confirm the striking influence of the
dressed pion on the antikaon in-medium spectrum
obtained in ref.~\cite{tolos02}. Possible reasons for this
discrepeancy can reside in the different pion spectral functions
and transition amplitudes coupling the antikaon-nucleon channel to
pion-hyperon channels. The pion spectral function \cite{ramos94}
used in
ref.~\cite{tolos02} shows very strong softening (more than
$100\,$MeV at $400\,$MeV momentum), which was not obtained in
self-consistent schemes \cite{xia94,korpa95,korpa03a} and the
relativistic delta-hole model \cite{lutz03}.

\begin{figure}[htb]
\vspace*{-0.2cm}
                 \insertplot{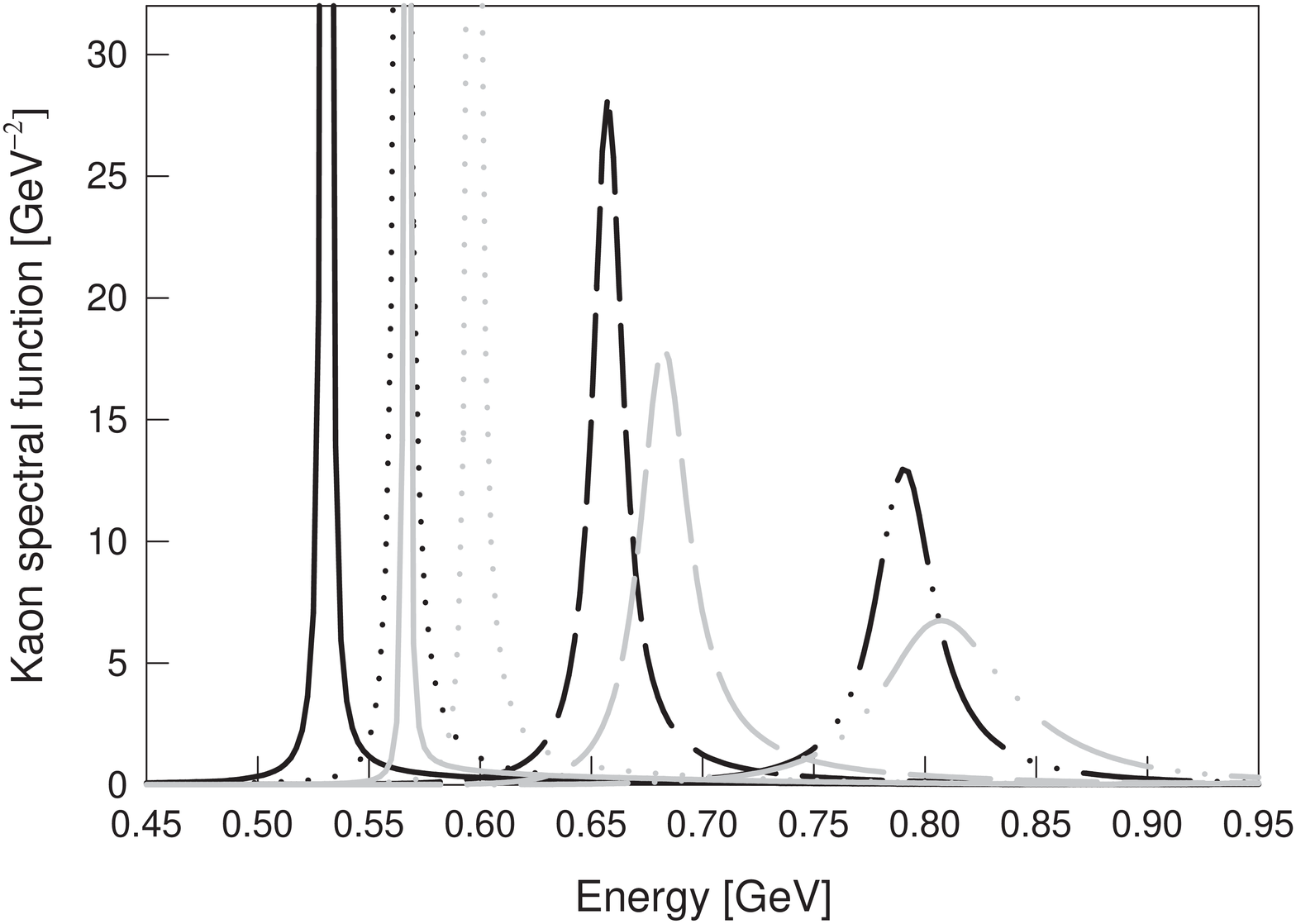}

\vspace*{-1.4cm}
\caption[]{The kaon spectral function for different momenta:
0 (solid line), $200\,$MeV (dotted line), $400\,$MeV (dashed line),
and $600\,$MeV (dash-dot-dot line). Black lines show the result
at saturation density, $\rho=0.17\,$fm$^{-3}$, gray lines at twice
saturation density, $\rho=0.34\,$fm$^{-3}$.
}
\label{fig3}
\end{figure}

We now turn to the kaon (${\rm{K}}^+, {\rm K}^0$) properties in
cold isospin-symmetric nuclear
medium. The computational procedure is identical to the one used for
the antikaons \cite{lutz02}, the difference being only the input
vacuum scattering amplitudes, which now refer to kaon-nucleon
scattering. They are again taken from ref.~\cite{lutz02a}, based
on an extensive fit to experimental data. Figure \ref{fig3} shows
the kaon spectral function at saturation density (black lines) and
twice the saturation density (gray lines), for momenta 0, $200\,$MeV,
$400\,$MeV and $600\,$MeV. We do not get any structure in the
spectral function, the medium effect shows up as a repulsive shift
of the quasiparticle peak, which also acquires a finite width.
The latter is quite small (less than $5\,$MeV) for momenta below
$200\,$MeV at saturation density, but increases to $15\,$MeV at
$400\,$MeV momentum. The positions of the peaks are a few MeV
below the value given by the $\sqrt{M_*^2+\vec p\,^2}$ for momenta
$|\vec p|>200\,$MeV, where $M_*$ is the peak's position for $|\vec p|=0$.

\begin{figure}[htb]
\vspace*{-0.2cm}
                 \insertplot{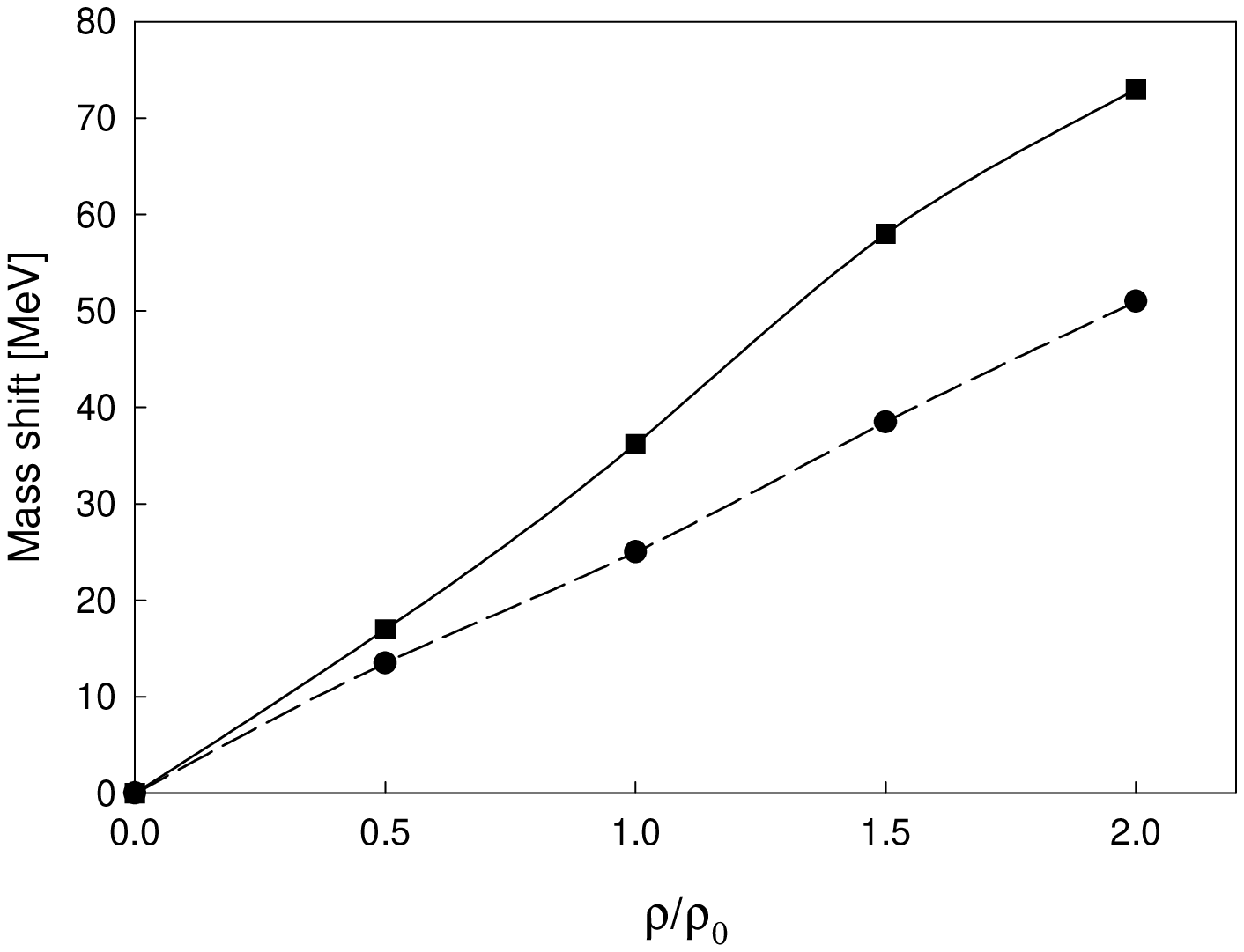}

\vspace*{-2.6cm}
\caption[]{The kaon mass shift as function of the nucleon
density, with $\rho_0$ denoting the saturation density.
The circles show the results based on the vacuum kaon-nucleon
scattering amplitude, while the squares correspond to the
self-consistent result. The lines smoothly connect the
points.
}
\label{fig4}
\end{figure}

In fig.~\ref{fig4} we show the kaon-mass shift as a function of nucleon
density. The circles show the mass shift obtained with vacuum scattering
amplitudes, while the squares correspond to the full self-consistent solution,
based on the in-medium scattering amplitudes. The lines are drawn simply as
smooth connections of points. The self-consistent mass shift exceeds the
one based on the vacuum scattering amplitude, but shows signs of beginning
saturation above $1.5\,\rho_0$, however the results should be regarded
with caution for densities reaching $2\,\rho_0$, since neglected nucleon
correlations may affect them.

\section{Conclusions}\label{concl}
We studied the properties of kaons and antikaons in cold nuclear
medium by solving self-consistently the in-medium Bethe-Salpeter
equation. For the antikaons we included the effect of the dressed
pion in the coupled pion-hyperon channels. The pion dressing did
not change the in-medium properties of hyperons and antikaons in a
significant way, introducing in general slightly more broadening
for the hyperons and smoothing somewhat the antikaon spectral
function at small momenta. The only exception is the
$\Sigma(1690)$ which was affected very little by the dressing of
the antikaon, but now suffers a shift of $-20\,$MeV and significant
broadening, its width increasing to about $130\,$MeV at zero momentum.
The minor
effect of the pion dressing is in contrast with the result of
ref.~\cite{tolos02} where a significant change of the antikaon
spectrum was reported, when using a pion spectral function with a
large shift of the main maximum towards smaller energy.

For the kaons (${\rm{K}}^+, {\rm K}^0$) we obtain a repulsive mass
shift of $36\,$MeV at saturation density and only a modest
broadening of the quasiparticle peak. The position of the peak is
a few MeV below the value of $\sqrt{M_*^2+\vec p\,^2}$ for momenta
$|\vec p|>200\,$MeV, where $M_*$ is the position for $|\vec p|=0$.
The self-consistently obtained mass shift is larger than the one
obtained from vacuum scattering amplitude, the difference increasing
from $11\,$MeV at saturation density to $22\,$MeV at twice the
saturation density.

\section*{Acknowledgement}
C.L.K would like to thank the ``Nederlandse Organisatie voor Wetenschappelijk
Onderzoek" (NWO) for providing a
visitors stipend and the Kernfysisch Versneller Instituut (Groningen)
for the kind hospitality.

\vfill\eject
\end{document}